\documentclass[12pt]{article}
\usepackage{fullpage, amsfonts, natbib}
\usepackage{algorithm, enumerate, color,url,multirow}
\usepackage[noend]{algpseudocode}
\usepackage{times}
\usepackage{dsfont}
\usepackage{graphicx}
\usepackage{amsmath}
\usepackage{caption}
\usepackage{subcaption}
\usepackage{mathrsfs}
\usepackage{amssymb}
\usepackage{pdfpages}
\usepackage{tabularx}
\usepackage{multirow}
\usepackage{gensymb}
\usepackage{amsthm}
\usepackage{relsize}
\usepackage[hidelinks]{hyperref}
\hypersetup{colorlinks=true,citecolor=blue}
\DeclareMathOperator*{\argmax}{argmax}
\providecommand{\keywords}[1]{\textbf{Keywords} #1}
\newcommand\ringring[1]{%
  {
   \mathop{\kern0pt #1}\limits^{
     \vbox to-1.85ex{
       \kern-2ex 
       \hbox to 0pt{\hss\normalfont\kern.1em \r{}\kern-.45em \r{}\hss}%
       \vss 
     }
   }
  }
}

\newtheorem{thm}{Theorem}[]

\newcounter{lastnote}

\title{High-dimensional Multivariate Mediation\\
 with Application to Neuroimaging Data}

\author{
Oliver Y. Ch\'en$^1$,  Ciprian M. Crainiceanu$^1$, Elizabeth L. Ogburn$^1$, \\ Brian S. Caffo$^1$, Tor D. Wager$^2$, Martin A. Lindquist$^1$ \\ \\
$^1$ Department of Biostatistics\\
\bigskip
Johns Hopkins Bloomberg School of Public Health \\ 
$^2$ Department of Psychology and Neuroscience\\
University of Colorado Boulder
}

\date{}

\def\E{\mathbb{E}}

\def\bE{\mathbf{E}}

\def\by{\mathbf{y}}
\def\bY{\mathbf{Y}}

\def\1{\mathbf{1}}

\def\beps{\boldsymbol{\epsilon}}
\def\bmu{\boldsymbol{\mu}}
\def\bX{\mathbf{X}}

\def\bSigma{\boldsymbol{\Sigma}}
\def\bEta{\boldsymbol{\eta}}

\def\bzero{\boldsymbol{0}}
\def\bV{\mathbf{V}}
\def\bI{\mathbf{I}}
\def\bJ{\mathbf{J}}
\def\b1{\mathbf{1}}

\def\bD{\mathbf{D}}

\def\bV{\mathbf{V}}

\def\bU{\mathbf{U}}
\def\bD{\mathbf{D}}
\def\bX{\mathbf{X}}
\def\bY{\mathbf{Y}}
\def\bx{\mathbf{x}}
\def\by{\mathbf{y}}

\def\bM{\mathbf{M}}
\def\bW{\mathbf{W}}
\def\bw{\mathbf{w}}
\def\bff{\mathbf{f}}
\def\btheta{\boldsymbol{\theta}}
\def\bTheta{\boldsymbol{\Theta}}

\newcommand\independent{\protect\mathpalette{\protect\independenT}{\perp}}
\def\independenT#1#2{\mathrel{\rlap{$#1#2$}\mkern2mu{#1#2}}}

\begin{document}


\baselineskip24pt


\maketitle 

\thispagestyle{empty}

\newpage
\pagenumbering{roman}
\begin{abstract}
Mediation analysis is an important tool in the behavioral sciences for investigating the role of intermediate variables that lie in the path between a treatment and an outcome variable. The influence of the intermediate variable on the outcome is often explored using a linear structural equation model (LSEM), with model coefficients interpreted as possible effects. While there has been significant research on the topic, little work has been done when the intermediate variable (mediator) is a high-dimensional vector.  In this work we introduce a novel method for identifying potential mediators in this setting called the directions of mediation (DMs).  DMs linearly combine potential mediators into a smaller number of orthogonal components, with components ranked by the proportion of the LSEM likelihood (assuming normally distributed errors) each accounts for. 
This method is well suited for cases when many potential mediators are measured. Examples of high-dimensional potential mediators are brain images composed of hundreds of thousands of voxels, genetic variation measured at millions of SNPs, or vectors of thousands of variables in large-scale epidemiological studies. We demonstrate the method using a functional magnetic resonance imaging (fMRI) study of thermal pain where we are interested in determining which brain locations mediate the relationship between the application of a thermal stimulus and self-reported pain.
\end{abstract}


\newpage
\pagenumbering{arabic}
\section{Introduction}

\begin{figure}[htbp] 
\begin{center}
\includegraphics[width=7cm]{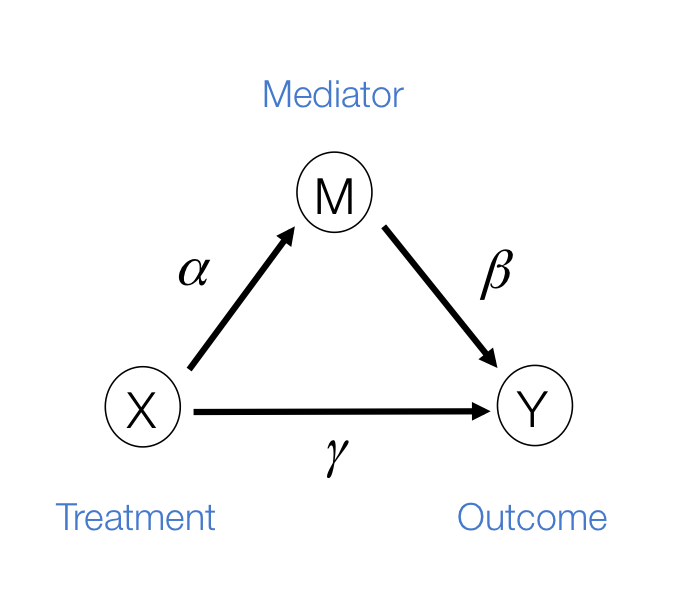}
\caption{The three-variable path diagram representing the standard mediation framework. The variables corresponding to $X$, $Y$, and $M$ are all scalars, as are the path coefficients $\alpha$, $\beta$, and $\gamma$.}
\label{fig:M_classical}
\end{center}
\end{figure}

Mediation and path analysis have been pervasive in the social and behavioral sciences (e.g., \cite{baron86}; \cite{mackinnon08}; \cite{preacher2008asymptotic}), and have found widespread use in many applications, including psychology, behavioral science, economics, decision-making, health psychology, epidemiology, and neuroscience.  In the past couple of decades the topic has also begun to receive a great deal of attention in the statistical literature, particularly in the area of causal inference (e.g., \cite{Holland88}; \cite{robins92}; \cite{Angrist96}; \cite{ten07}; \cite{albert08}; \cite{jo08}; \cite{sobel08}; \cite{vanderweele09}; \cite{imai10}; \cite{lindquist12}; \cite{pearl2014interpretation}).  When the effect of a treatment $X$ on an outcome $Y$ is at least partially directed through an intervening variable $M$, then $M$ is said to be a mediator. The three-variable path diagram shown in Figure \textcolor{blue} {\ref{fig:M_classical}} illustrates this relationship. The influence of the intermediate variable on the outcome is frequently ascertained using linear structural equation models (LSEMs), with the model coefficients interpreted as causal effects; see below for discussion of the assumptions under which this interpretation is warranted. Typically, interest centers on parsing the effects of the treatment on the outcome into separable direct and indirect effects, representing the influence of $X$ on $Y$ unmediated and mediated by $M$, respectively.

To date most research in mediation analysis has been devoted to the case of a single mediator, with some attention given to the case of multiple mediators (e.g., \cite{preacher2008asymptotic,vanderweele2013mediation}). However, high dimensional mediation has received scarce attention.  
Recent years have seen a tremendous increase of new applications measuring massive numbers of variables, including brain imaging, genetics, epidemiology, and public health studies. It has therefore become increasingly important to develop methods to deal with mediation in the high-dimensional setting, i.e., when the number of mediators is much larger than the number of observations. Such an extension is the focus of this work.
It is important to emphasize that even though we focus on high dimensional mediators in the context of LSEMs, the principles extend to any other model-based approach to mediation.

As a motivating example, consider functional magnetic resonance imaging (fMRI), which is an imaging modality that allows researchers to measure changes in blood flow and oxygenation in the brain in response to neuronal activation (\cite{ogawa1990brain}; \cite{kwong1992dynamic}; \cite{lindquist08}). In fMRI experiments, a multivariate time series of three dimensional brain volumes are obtained for each subject, where each volume consists of hundreds of thousands of equally sized volume elements (voxels). A number of previous studies have used fMRI to investigate the relationship between painful heat and self-reported pain (\cite{apkarian2005human}; \cite{bushnell2013cognitive}). Recently, studies have focused on trial-by-trial modeling of the relationship between the intensity of noxious heat and self-reported pain (\cite{wager13}; \cite{atlas2014brain}). In \cite{woo15}, for example, a series of thermal stimuli were applied at various temperatures (ranging from $44.3-49.3$ $\degree$C in $1$ $\degree$ increments) to the left forearm of each of $33$ subjects. In response, subjects gave subjective pain ratings at a specific time point following the offset of the stimulus. During the course of the experiment, brain activity in response to the thermal stimuli was measured across the entire brain using fMRI. One of the goals of the study was to search for brain regions whose activity level act as potential mediators of the relationship between temperature and pain rating.

In this context, we are interested in whether the effect of temperature, $X$, on reported pain, $Y$, is mediated by the brain response, $\bM$. Here both $X$ and $Y$ are scalars, while $\bM$ is the estimated brain activity measured over a large number of different voxels/regions. We assume that the values of $\bM$ are either parameters or contrasts (linear combinations of parameters) obtained by fitting the general linear model (GLM), where for each subject, the relationship between the stimuli and the BOLD response is analyzed at the voxel level \citep{lindquist2012estimating}. Standard mediation techniques are applicable to univariate mediators.  An early approach to mediation in neuroimaging \citep{caffo2008brain} took the  route of re-expressing the multivariate images into targeted, simpler, composite summaries on which mediation analysis was performed. In contrast, the identification of univariate mediators on a voxel-wise basis has come to be known as Mediation Effect Parametric Mapping (\cite{Wager08}; \cite{Wager09a}; \cite{Wager09b}) in the neuroimaging field. This approach, however, ignores the relationship between voxels, and identifies a series of univariate mediators rather than an optimized, multivariate linear combination.  A multivariate extension should focus on identifying latent brain components that may be maximally effective as mediators, i.e. those that are simultaneously most predictive of the outcome and predicted by the treatment.

Thus, in this work we consider the same simple three-variable path diagram depicted in Figure \textcolor{blue} {\ref{fig:M_classical}}, with the novel feature that the scalar potential mediator is replaced by a very high dimensional vector of potential mediators $\bM= ( M^{(1)}, M^{(2)}, \ldots M^{(p)} )^\intercal \in \mathbb{R}^p$.  While an LSEM can be used to estimate mediation effects (defined precisely below), in this setting there are too many mediators to allow reasonable interpretation (unless the model coefficients are highly structured) and there are many more mediators than subjects, precluding estimation using standard procedures. To overcome these problems, a new model, called the directions of mediation (DM) is developed. DM's linearly combine activity in different voxels into a smaller number of orthogonal components, with components ranked by the proportion of the LSEM likelihood (assuming normally distributed errors) each accounts for. Ideally, the components form a small number of uncorrelated mediators that represent interpretable networks of voxels. The approach shares some similarities with partial least squares (PLS) (\cite{wold1982soft}; \cite{wold1985partial}; \cite{krishnan2011partial}), which is a dimension reduction approach based on the correlation between a response variable (e.g. $Y$) and a set of explanatory variables (e.g. ${\bM}$). In contrast, for DM the dimension reduction is based on the complete $X$-${\bM}$-$Y$ relationship. 

This article is organized as follows. In Section \textcolor{blue} {\ref{MCMM}} we define direct and indirect effects for the multiple mediator setting. In Section \textcolor{blue} {\ref{sec:DM}} we introduce the directions of mediation, and provide an estimation algorithm for estimating the DM and its associated path coefficients when the mediator is high dimensional. In Section \textcolor{blue} {\ref{sec: sec4}} we discuss a method for performing inference on the DM. Finally, in Sections \textcolor{blue} {\ref{sec: sec5}} - \textcolor{blue} {\ref{sec: sec6}} the efficacy of the approach is illustrated through simulations and an application to the fMRI study of thermal pain.

\section{A Mutivariate Causal Mediation Model} \label{MCMM}

Let $X$ denote an exposure/treatment for a given subject (e.g., thermal pain),  and $Y$ an outcome (e.g.,  reported pain).  Suppose there are multiple mediators  ${\bf M} = (M^{(1)}, \cdots M^{(p)})$ in the path between treatment and outcome; in the fMRI context, the mediators are $p$ dependent activations over the $p$ voxels.  
Here we assume for simplicity that each subject is scanned under one condition.

Using potential outcomes notation (\cite{Rubin74}), 
let ${\bf M} (x)$ denote the value of the mediators if treatment $X$ is set to $x$. Similarly, let $Y (x, {\bf m})$ denote the
outcome if $X$ is set to $x$ and ${\bf M}$ is set to ${\bf m}$. 
The controlled unit direct effect of $x$ vs. $x^{*}$ is defined as $Y(x,{\bf m}) - Y(x^*, {\bf m})$, the natural unit direct effect as $Y(x,{\bf M}(x^*)) - Y(x^*, {\bf M}(x^*))$, and the 
natural unit indirect effect as $Y(x,{\bf M}(x)) - Y(x, {\bf M}(x^*))$.  Note that for these nested counterfactuals to be well-defined it must be hypothetically possible to intervene on the mediator without affecting the treatment. 

The total unit effect is the sum of the natural unit direct and unit indirect effects, i.e.
\begin{equation}
Y(x,{\bf M}(x)) - Y(x^*, {\bf M}(x^*)) = Y(x,{\bf M}(x)) - Y(x, {\bf M}(x^*)) + Y(x,{\bf M}(x^*)) - Y(x^*, {\bf M}(x^*))
\label{eq:total}
\end{equation}
Note that the direct effect could also be defined as $Y(x,{\bf M}(x)) - Y(x^*, {\bf M}(x))$.  In general, this would lead to a different decomposition of the total effect;
however, as we consider linear models below, this is not of further concern.
Suppose the following four assumptions hold for the set of mediators:
\begin{eqnarray}
Y(x,{\bf M}(x)) \independent X \nonumber  \\
Y(x,m) \independent {\bM} | X  \nonumber \\
{\bf M}(x)  \independent X \nonumber \\
Y(x,m) \independent {\bM}(x^*). 
\label{eq:mediation}
\end{eqnarray}
In words, these assumptions imply there is no confounding for the relationship between: (i) treatment $X$ and outcome $Y$; (ii) mediators $\bM$ and outcome $Y$; (iii) treatment $X$ and mediators $\bM$; and (iv) no confounding for the relationship between mediator and outcome that is affected by the treatment. 
See \cite{robins2010alternative} and \cite{pearl2014interpretation} for detailed discussion of these assumptions, and for a critical evaluation of these assumptions in the high-dimensional setting see \cite{huang2015hypothesis}. \cite{vanderweele2013mediation} showed that under (\ref{eq:mediation}) the average direct and indirect effects are identified from the regression function for the observed data.  Suppose then
(\ref{eq:mediation}) and the following model for
the observed data hold: 
\begin{eqnarray}
E(M^{(j)}|X =x) &=& \alpha_0+ \alpha_{j} x  \hspace{1cm}  \mbox{for \hspace{2mm} $j = 1, \ldots, p$ \nonumber }\\
E(Y|X = x, {\bf M} = {\bf m}) &=& \beta_0+ \gamma x+  \beta_{1}M^{(1)} + \beta_{2}M^{(2)} + \cdots + \beta_{p}M^{(p)}. \label{RegY}
\end{eqnarray}
Note that this model encodes the assumptions of linear relations among treatment, mediators, and outcome and, importantly, the absence of any treatment-mediator interaction in the outcome regression.  When the treatment interacts with one or more of the mediators, the LSEM framework considered in this paper is not appropriate for mediation analysis \citep{ogburn2012commentary}. 

The average controlled direct effect, average natural direct effect and average indirect effect are expressed as follows:
\begin{eqnarray}
E(Y(x,{\bf m}) - Y(x^*, {\bf m})) &=& \gamma(x-x^{*}) \\
E(Y(x,{\bf M}(x^*)) - Y(x^*, {\bf M}(x^*))) &=& \gamma(x-x^{*}) \label{ANDE} \\
E(Y(x,{\bf M}(x)) - Y(x, {\bf M}(x^*))) &=& (x-x^{*}) \sum_{j=1}^p {\alpha_{j} \beta_{j}}. \label{AIDE}
\end{eqnarray}
Note the average controlled direct effect and natural direct effect are equivalent whenever there is no treatment-mediator interaction, as is assumed throughout.

When the counterfactuals are well-defined and the assumptions in (\ref{eq:mediation}) hold, the right hand sides of (\ref{ANDE}) and (\ref{AIDE}) identify causal mediation effects.  When one or more of the assumptions in (\ref{eq:mediation}) fail to hold, or if the counterfactuals are not well-defined, the right hand sides of (\ref{ANDE}) and (\ref{AIDE}) may still be used in exploratory analysis to help identify potential mediators. For example, they could identify linear combinations of voxels that correspond to specific brain functions, suggesting mediation through correlates of those brain functions. Throughout, for simplicity, we use ``direct effect" and ``indirect effect" to refer to the right hand sides of (\ref{ANDE}) and (\ref{AIDE}), respectively; we are agnostic throughout as to whether these expressions can be interpreted causally or should be taken as exploratory.  Similarly, we use ``mediator" agnostically to refer to variables that temporally follow treatment and precede outcome and potentially may lie on a causal pathway between them.

\begin{figure}[htbp] 
\begin{center}
\includegraphics[width=15cm]{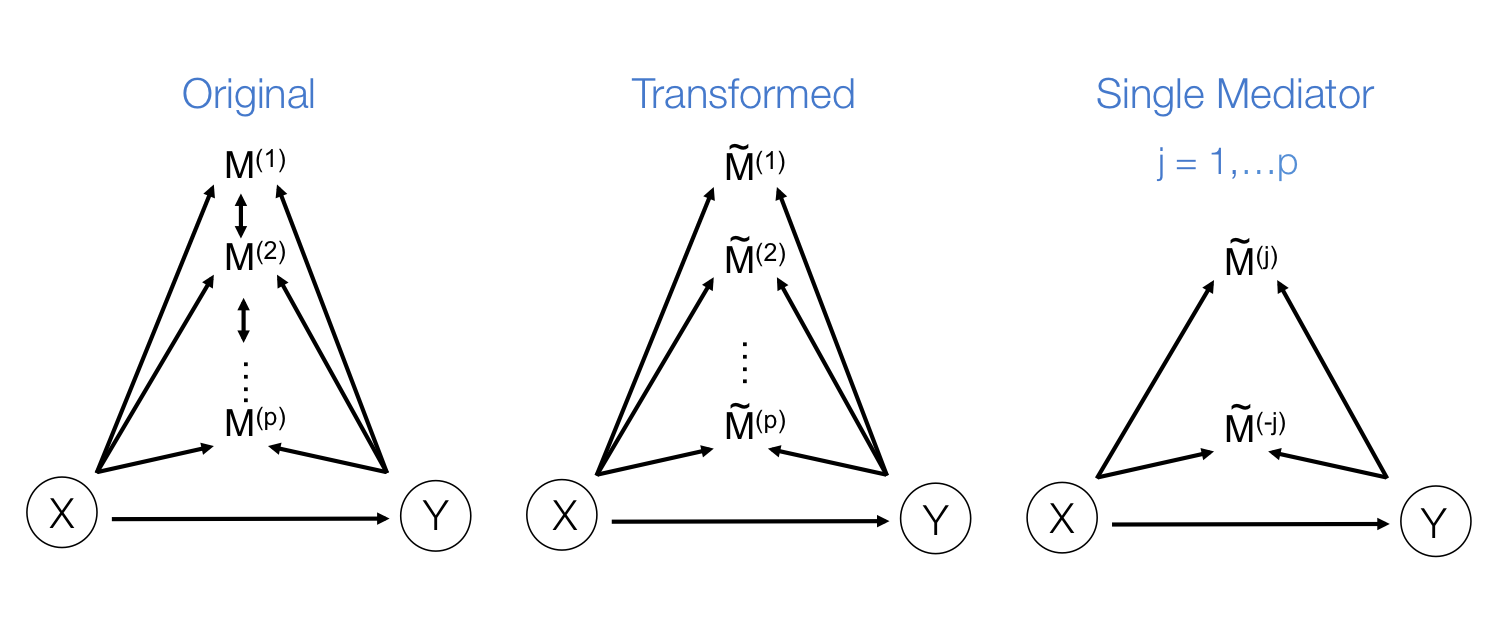}
\caption{(Left) The three-variable path diagram used to represent multivariate mediation. Here the $p$ mediators are assumed to be correlated. (Center) A similar path diagram after an orthogonal transformation of the mediators. Now the $p$ mediators are independent of one another, allowing for the use of a series of LSEMs (Right), one for each transformed mediator, to estimate direct and indirect effects.}
\label{fig:fig2}
\end{center}
\end{figure}

Fitting the system (\ref{RegY}) is straightforward if the number of mediators is small. However, the estimates become unstable as $p$ increases, and in fMRI 
the number of mediators will greatly exceed the sample size. Therefore we seek an orthogonal transformation of the mediators.
This both simplifies and stabilizes the parameter estimates in the model (\ref{RegY}), allowing us to estimate the 
direct and indirect effects using a series of LSEMs, one for each transformed mediator; see Fig. \ref{fig:fig2} for an illustration.  The novelty of our approach lies in choosing the transformation so that the transformed mediators are ranked by the proportion of the likelihood of the full LSEM that they account for. 
This has the benefit of potentially: (i) providing more interpretable mediators (i.e. linear combinations of voxels rather then individual voxels); and (ii)  reducing the number of mediators needed to estimate the indirect effect.

\section{Directions of Mediation} \label{sec:DM}

In this section we introduce a transformation of the space of mediators, determined by finding linear combinations of the original mediators that (i) are orthogonal; and (ii) are chosen to maximize the likelihood of the underlying three-variable SEM. 
We first formulate the model before introducing an estimation algorithm.  We conclude with a discussion regarding estimation for the case when $p >> n$.
 
\subsection{Model Formulation} \label{sec: sec3.2}
Let $X_i$ and $Y_i$ denote univariate variables, and $\bM_i = (M_{i}^{(1)}, M_{i}^{(2)}, \ldots M_{i}^{(p)} )^\intercal  \in \mathbb{R}^p$, for $i = 1, \ldots, n$.
We denote the full dataset $\Delta = (\bx, \by, \bM)$, where $\bx = (X_1, \ldots X_n)^\intercal \in \mathbb{R}^n$, $\by = (Y_1, \ldots Y_n)^\intercal \in \mathbb{R}^n$, and $\bM = (\bM_1, \ldots \bM_n)^\intercal \in \mathbb{R}^{n \times p}$.  
Now let $\bW = (\bw_1, \bw_2, \dots \bw_q) \in \mathbb{R}^{p \times q}$ be a linear transformation matrix, where $\bw_d = (w_d^{(1)}, w_d^{(2)}, \ldots w_d^{(p)})^\intercal \in \mathbb{R}^p$, for $d = 1, \ldots, q$; and let ${\tilde \bM} = \bM \bW = ({\tilde \bM}_1, {\tilde \bM}_2, \ldots {\tilde \bM}_n)^\intercal$ where $ {\tilde \bM}_i = \bM_i^\intercal \bW = ({\tilde M}_i^{(1)}, \ldots,  {\tilde M}_i^{(d)}, \ldots {\tilde M}_i^{(q)})^\intercal$ with ${\tilde M}_i^{(d)} = \bM_i^\intercal \bw_d = \sum_{k=1}^{p} M_i^{(k)}w_d^{(k)}$. We assume the relationship between the variables is given by the following LSEM:
\begin{eqnarray} 
{\tilde M}_i^{(j)} &=& \alpha_0 + \alpha_j X_i + \epsilon_i  \label{eq:eq1} \hspace{1cm}  \mbox{for \hspace{2mm} $j = 1, \ldots, q$} \nonumber \\
Y_i &=& \beta_0 + \gamma X_i + \beta_1 {\tilde M}_i^{(1)} + \beta_2 {\tilde M}_i^{(2)} + \ldots + + \beta_p {\tilde M}_i^{(q)} + \xi_i   \label{eq:eq2}
\end{eqnarray}
where $\epsilon_{i}$ and $\xi_{i}$ are i.i.d. bivariate normal with mean $0$ and variances $\sigma_\epsilon^2 $ and $\sigma_\xi^2 $.
The parameters of the LSEM can be estimated using linear regression. However, under the additional condition that the new transformed variables ${\tilde M}^{(j)}$ are orthogonal, we can estimate the parameters  separately for each ${\tilde M}^{(j)}$. 
Thus, for each $j = 1, \ldots, q$ we can fit the following LSEM:
\begin{eqnarray} 
{\tilde M}_i^{(j)} &=& \alpha_0 + \alpha_j X_i + \epsilon_i   \nonumber \\
Y_i &=& \beta_0 + \gamma X_i + \beta_j {\tilde M}_i^{(j)} + \eta_i   \label{eq:eq2:uni}
\end{eqnarray}
where $\epsilon_i \sim N (0, \sigma_\epsilon^2 )$ and $\eta_i \sim N (0, \sigma_\eta^2 )$, for $i = 1, \ldots, n$.

Let $\btheta:= (\alpha_0, \alpha_1, \beta_0, \beta_1, \gamma) \in \mathbb{R}^5$ be the parameter vector for the LSEM in 
\eqref{eq:eq2:uni} for $j=1$. We seek
to simultaneously estimate $\btheta$ and find the 
first direction of mediation (DM) $\bw_1$, defined as the linear combination of the elements of $\bM$ that maximizes the likelihood of the underlying LSEM.
In our motivating example, $\bw_1$ is a linear combination of the voxel activations. Thus, similar to principal components analysis (PCA) (\cite{andersen1999principal}) or independent components analysis (ICA) (\cite{mckeown1997analysis,calhoun2001method}) when applied to fMRI data, the weights can be mapped back onto the brain, with the resulting maps interpreted as coherent networks that together act as mediators of the relationship between treatment and outcome. Also like PCA, subsequent directions can be found that maximize the likelihood of the model, conditional on these being orthogonal to the previous directions.

To formalize, let $\mathscr{L}(\Delta; \bw_1, \boldsymbol { \theta})$ be the joint likelihood of the SEM stated in \eqref{RegY}. The \textit{Directions of Mediation} are defined as follows:

\bigskip
\noindent \textit{Step 1}: The $1^{st}$ DM is the vector $\bw_1 \in \mathbb{R}^p$, with norm 1, that maximizes the conditional joint likelihood $\mathscr{L}(\Delta, \boldsymbol { \theta}; \bw_1)$, i.e.
\begin{equation*}
\hat{\bw}_1 |\boldsymbol { \theta} = \argmax
\bigg \{ 
\mathscr{L}(\Delta, \boldsymbol { \theta}; \bw_1)
\bigg \},
\end{equation*} 
subject to
\begin{equation*}
\big \{ \bw_1 \in \mathbb{R}^p: \left \Vert \bw_1 \right \Vert_2 =1 \big \}.
\end{equation*}
\bigskip
\noindent \textit{Step 2}: The $2^{nd}$ DM is the vector $\bw_2 \in \mathbb{R}^p$, with norm 1 and orthogonal to $\bw_1$, that maximizes the conditional joint likelihood $\mathscr{L}(\Delta, \boldsymbol { \theta}, \bw_1; \bw_2)$, i.e.
\begin{equation*}
\hat{\bw}_2 |\boldsymbol { \theta},\bw_1 = \argmax 
\bigg \{ 
\mathscr{L}(\Delta, \boldsymbol { \theta}, \bw_1; \bw )
\bigg \}
\end{equation*} 
subject to 
\begin{equation*}
\bigg \{ \bw_2 \in \mathbb{R}^p: \left \Vert \bw_2 \right \Vert_2 =1,
\bw_1 \bw_2^\intercal=0 \bigg \}.
\end{equation*}
\begin{center}
$\vdots$
\end{center}

\noindent \textit{Step k}: The $k^{th}$ DM is the vector $ \bw_k$, with norm 1 and orthogonal to $\bw_1, \ldots, \bw_{k-1}$, that maximizes the conditional joint likelihood $\mathscr{L}(\Delta, \bw_1, \ldots, \bw_{k-1};\bw_k)$, i.e.
\begin{equation*}
\hat{\bw}_k |\boldsymbol { \theta},\bw_1, \ldots,\bw_{k-1}
= \argmax
\bigg \{ 
\mathscr{L}(\Delta, \boldsymbol { \theta}, \bw_1, \ldots, \bw_{k-1};\bw)
\bigg \} 
\end{equation*} 
subject to 
\begin{equation*}
\bigg \{  \bw_k \in \mathbb{R}^p: \left \Vert \bw_k \right \Vert_2 =1, 
\bw_{k'}\bw_k^\intercal=0, \forall k' \in \{1, \ldots, k-1 \}
\bigg \}.
\end{equation*}

\bigskip
\noindent {\it Remark:} According to the model formulation the signs of the DMs are unidentifiable. 

\subsection{Estimation} \label{sec: sec3.3}

Here we describe how to estimate the parameters associated with the first DM. Assuming joint normality, the joint $\log$ likelihood function for $\bw_1$ and $\boldsymbol { \theta}$, $\mathscr{L}(\Delta; \bw_1, \boldsymbol { \theta})$, can be expressed as:
\begin{equation} \label{eq:eq9}
\mathscr{L}(\Delta; \bw_1, \btheta)  \propto  g_1(\Delta; \bw_1, \btheta),
\end{equation}
where    
$g_1(\Delta; \bw_1, \btheta) \equiv
- \big \{
\frac{1}{\sigma_{\epsilon}^2} 
\left \Vert \by - \beta_{0}-\bx\gamma_1 - \bM \bw_1 \beta_1 
\right \Vert_2
+ \frac{1}{\sigma_{\eta}^2}
\left \Vert
\bM \bw_1-\alpha_{0} - \bx \alpha_1
\right \Vert_2
\big \}$.\\
The goal is to find both the parameters of the LSEM and the first DM that jointly maximize $g_1(\Delta; \bw_1, \boldsymbol { \theta})$, under the constraint that the $L_2$ norm of $\bw_1$  equals 1. 
Consider the Lagrangian
\begin{equation*}
L(\Delta; \bw_1, \btheta, \lambda) = g_1(\Delta; \bw_1, \btheta) + \lambda ( \left \Vert \bw_1 \right \Vert_2 - 1).
\end{equation*} 
The dual problem can be expressed:
\begin{equation*}
(\hat{\bw}_1, \hat{\boldsymbol { \theta}}) | \lambda = \argmax _{\big \{
\substack{
          \bw_1 \in \mathbb{R}^p\\
 \btheta \in \mathbb{R} ^5            
           } 
\big \} }
 L(\Delta; \bw_1, \btheta, \lambda)
\end{equation*}
where $\lambda$ is the Lagrange multiplier.  To solve this problem we propose a method where $\lambda$ is profiled out by one set of parameters of interest. We  establish, under the assumption that
the first partial derivatives of the objective function and the constraint function exist, the closed form solution for the path coefficients, the first DM, and $\lambda$ as follows: 
\begin{eqnarray} \label{eq:eq11}
\hat{\bw}_1 | \boldsymbol{\theta}, \lambda &=& f_1 (\Delta; \lambda, \btheta)\\
\hat{ \lambda} | \boldsymbol{\theta} &=& \arg _{\lambda \in \mathbb{R}^1} 
\bigg \{ f_2 (\Delta; \lambda, \btheta) =1 
\bigg \}  \label{eq:eq12} \\
\hat{\btheta}|\hat{\bw}_1,\hat{\lambda} &=& \argmax_{\btheta \in \mathbb{R}^5} L(\Delta; \hat{\bw}_1, \btheta, \hat{\lambda}) \label{eq:eq12b}
\end{eqnarray}
where $f_1 (\Delta; \lambda, \btheta) = (\lambda \bI + \boldsymbol{\psi} (\boldsymbol{\theta}) )^{-1} \boldsymbol{\phi} (\boldsymbol{\theta}) $; $f_2 (\Delta; \lambda, \btheta) = \left \Vert 
(\lambda \bI + \boldsymbol{\psi} (\btheta) )^{-1} \boldsymbol{\phi} (\btheta)
\right \Vert_2$, $\boldsymbol{\psi} (\boldsymbol{\theta}) = \bM^\intercal\bM\beta_1^2 / \sigma_{\epsilon_1}^2 + \bM^\intercal \bM / \sigma_{\eta_1}^2$,
and 
$
\boldsymbol{\phi} (\btheta) = \bM^\intercal(\alpha_{0}+\alpha_1 \bx ) / \sigma_{\eta_1}^2  + \bM^\intercal(\by - \beta_{0} - \bx \gamma_1)\beta_1 / \sigma_{\epsilon_1}^2$.
Using these results we outline an iterative procedure for jointly estimating the first direction of mediation and path parameters as described in Algorithm \ref{DM}.
Further, in the Supplemental Material we show that the estimated parameters are consistent and asymptotically normal (see Theorems 1 and 2).

\bigskip
\begin{algorithm}
\caption{First DM}
\label{DM}
\begin{algorithmic}[h]
\State \textbf{Step 0:}
Initiate $\boldsymbol{\theta}$, denoted $\boldsymbol{\theta}_1^{(0)}$.
\State \textbf{Step 1:}
For each $k$, set: 
\begin{eqnarray} \label{eq:eq15}
\hat{\lambda}^{(k)} | \boldsymbol{\theta}_1^{(k)} &=& {\arg}_{\lambda \in \mathbb{R}^1} \bigg \{ 
f_2(\Delta; \lambda, \btheta_1^{(k)}) =1 
\bigg \} \\
\hat{\bw}_1^{(k)}|\boldsymbol{\theta}_1^{(k)}, \hat{\lambda}^{(k)} &=& 
f_1(\Delta; \hat{\lambda}^{(k)}, \btheta_1^{(k)})
 \label{eq:eq16} \\
\hat{\boldsymbol{\theta}}_1^{(k + 1)}|\hat{\bw}_1^{(k)}, \hat{\lambda}^{(k)} &=& {\arg \max}_{\btheta_1 \in \mathbb{R}^5 } \bigg \{ L( \Delta; \hat{\bw}_1^{(k)}, \boldsymbol{\theta}_1^{(k)}, \hat{\lambda}^{(k)})
\bigg \}. \label{eq:eq17}
\end{eqnarray} 
\State \textbf{Step 2:}
Repeat Step 1 until convergence; each time set $k=k+1$.
\end{algorithmic}
\end{algorithm}


\subsection{Higher Order Directions of Mediation} \label{HODM}

To estimate higher order DMs we investigated two alternative approaches. The first uses additional penalty parameters (one for each additional constraint), and the second subtraction and \textit{Gram-Schmidt} projections. While the former approach is likely to achieve global maxima, the latter is computationally more efficient, and provides a good approximation of higher order DMs; thus we focus on this approach here.
Using this approach, estimates of  the $k^{\text{th}}$ direction of mediation, $\hat{ \textbf{w}}_k$, and the associated path coefficients, $\hat{\boldsymbol { \theta}}_k$, are obtained by computing:
\begin{equation*}
(\hat{ \textbf{w}}_k, \hat{\boldsymbol { \theta}}_k ) | \lambda= \argmax
\bigg \{   g_k(\Delta, \hat{ \textbf{w}}_1, \ldots,\hat{ \textbf{w}}_{k-1}  ; \textbf{w}_k, \boldsymbol { \theta}_k )
- \lambda \big (
\left\Vert  \bw_k (\bx) \right\Vert_2
 - 1  \big )  
 \bigg \},
\end{equation*}
subject to
\begin{equation*}
\bigg \{  
\btheta_k \in \mathbb{R}^{k+4}, \textbf{x} \in \bar{\mathbb{R}} ^p: \bw_k (\bx): =  \bx - \sum_{i=1} ^{k-1} \text{Proj}_{\hat{\bw}_i } (\bx) 
\btheta_k  \bigg \}
\end{equation*}
where $ \text{Proj}_{\hat{ \textbf{w}}_i } (\textbf{x}) = 
\dfrac{ \langle \textbf{x}, \hat{ \textbf{w}}_i \rangle } { \langle \hat{ \textbf{w}}_i, \hat{ \textbf{w}}_i \rangle} \hat{ \textbf{w}}_i$, $\forall i \in \{ 1, \ldots, k-1 \}$.
The performance of the projection approach is evaluated through extensive simulations in Section \ref{sec: sec5}.

\subsection{High-dimensional Directions of Mediation} \label{sec: sec3.5}

The estimation procedure described in \ref{sec: sec3.3} works well in the low-dimensional setting, but becomes cumbersome as $p$ increases.  Therefore it is critical to augment it with a matrix decomposition technique. Here we use a generalized version of Population Value Decomposition (PVD) \citep{caffo2010two, crainiceanu11}, which in contrast to Singular Value Decomposition (SVD) provides population-level information about $\bM$. We begin by introducing the generalized version of PVD and thereafter illustrate its use in estimating the DMs.  Throughout we assume that the data for each subject $i$ is stored in an $T_i \times p$ matrix, $\bM_i$, whose $j^{\text{th}}$ row contains voxel-wise activity for the measurements of the $j^{\text{th}}$ trail for the $i^{\text{th}}$ subject. All $\bM_i$ matrices are stacked vertically to form the $n \times p$ matrix $\bM$, where $n = \sum_{i=1} ^N T_i$. 

\subsubsection{Generalized PVD}
The PVD framework assumes that the number of trials per subject is equal, which is not the case in many practical settings. To address this issue, we introduce Generalized Population Value Decomposition (GPVD), which allows the number of trials per subject to differ, while maintaining the dimension reduction benefits of the original. The GPVD of $\bM_i$ is given by
\begin{equation} \label{eq:eq28}
\bM_i = \bU_i^B \tilde{\bV}_i \bD + \bE_i,
\end{equation} 
where $\bU_i^B$ is an $T_i \times B$ matrix, $\tilde{\bV}_i$ is an $B \times B$ matrix of subject-specific coefficients, $\bD$ is a $B \times p$ population-specific matrix, $\bE_i$ is an $T_i \times p$ matrix of residuals. Here $B$ is chosen based upon a criteria such as total variance explained.

Below we introduce a step-by-step procedure for obtaining the GPVD. \medskip \\
\noindent \textbf{Step 1:}  
For each subject $i$, use SVD to compute: 
$\bM_i = \bU_i \bSigma_i \bV^\intercal_i \approx  \bU^B_i \bSigma^B_i (\bV^B_i)^\intercal$
where $\bU^B_i$ consists of the first $B$ columns of $\bU_i$, $\bSigma^B_i$ consists of the first $B$ diagonal elements of $\bSigma_i$, and $\bV^B_i$ consists of first $B$ columns of $\bV_i$. 

\medskip
\noindent \textbf{Step 2:} Form the $p \times NB$ matrix $\bV:= [\bV^B_1 , \ldots,  \bV^B_N ]$. When $p$ is reasonably small, use SVD to compute the eigenvectors of $\bV$. The $p \times B$ matrix $\bD$ is obtained using the first $B$ eigenvectors. When $p$ is large, performing SVD is computationally impractical due to memory limitations. Here instead perform a block-wise SVD \citep{zipunnikov11}, and compute the matrix $\bD$ as before. Here  $\bD$ contains common features across subjects. At the population level $\bV \approx \bD(\bD^\intercal \bV)$, and at the subject level $\bV^B_i \approx \bD(\bD^\intercal \bV^B_i)$. 

\medskip
\noindent \textbf{Step 3:} 
\noindent The GPVD in \textcolor{blue}{\eqref{eq:eq28}} can be summarized as follows:
\begin{equation} \label{eq:eq29}
\begin{split}
\bM_i &= \bU_i \bSigma_i \bV^\intercal _i  \approx  \bU^B_i \bSigma^B_i (\bV^B_i)^\intercal\\
& \approx \bU^B_i \underbrace{\{ \bSigma^B_i (\bV^B_i)^\intercal \bD^T \}}_{\tilde{V}_i} \bD = \bU^B_i \tilde{\bV}_i \bD,
\end{split}
\end{equation}
where $\bU^B_i $, $\bSigma^B_i$, and $\bV^B_i$ are obtained from Step 1, and $\bD$ from Step 2. The first approximation in \textcolor{blue} {\eqref{eq:eq29}} is obtained by retaining the eigenvectors that explain most of the observed variability at the subject level. The second results from projecting the subject-specific right eigenvectors on the corresponding population-specific eigenvectors.

\subsubsection{Estimation using GPVD}
To estimate the DMs, perform GPVD on $\bM = 
[ \bM_1^\intercal, \cdots, \bM_n^\intercal ] ^\intercal 
=
\big [ (\textbf{U}_1 \tilde{\bV}_1 \bD)^\intercal, \cdots, (\textbf{U}_n \tilde{\bV}_n \bD)^\intercal \big ]^\intercal$. Next, stack all $T_i \times B $ matrices $ \bU_i \tilde{\bV}_i$ vertically to form an $n \times B$ matrix
\begin{equation} \label{eq:M_tilde}
\breve{\bM} = 
\big[ 
(\bU_1 \tilde{V}_1)^\intercal, \cdots, (\bU_n \tilde{V}_n)^\intercal
\big ]^\intercal
\end{equation}
Let $\breve{\bw} = \bD \bw$, where $\breve{\bw}$ is $B \times 1$. 
Finally, place $\breve{\bM}$ and $\breve{\bw}$ into (\ref{eq:eq2}).
Since $\bD$ can be obtained via GPVD, we can retrieve the original estimator of the high dimensional direction of mediation, $\hat{\bw}$, via the generalized inverse, i.e., 
\begin{equation} \label{eq: ginverse}
\hat{\bw} = \bD ^{-} \breve{\bw}^{est}
\end{equation} 
where $\breve{\bw}^{est}$ is the estimated $\breve{\bw}$ and $ ^- $ indicates the generalized inverse.

\section{Inference} \label{sec: sec4}
In low-dimensional settings, we can obtain variance estimates for the first DM and the path coefficients using \textit{Theorems 1} and \textit{2} from the Supplemental material. In high dimensional settings, variance estimation using the generalized inverse is under-estimated since the $\bD$ obtained from \textcolor{blue} {\eqref{eq:eq29}} is random. Even if we were to adjust for this, the covariance estimation of $\bD$ ($B \times p, B \ll p$) is computationally infeasible. Therefore, using the bootstrap to perform inference is a natural alternative. 

Consider $\bM = \breve{\bM} \bD$, where $\bM$ is $n \times p$, $\breve{\bM}$ is $n \times B$, $\bD$ is $B \times p$, and $B < n \ll p$.  The bootstrap procedure can be outlined as follows:
\begin{enumerate}
\item Bootstrap $n$ rows from $\breve{\bM}$, stack them horizontally and form the $n \times B$ matrix $\breve{\bM}_{ (j)}$;
\item Obtain $\hat{\breve{ \bw }}_{ (j) }$ from $\breve{ \textbf{ M }}_{(j)}$, where $\hat{\breve{\bw}}_{ (j) }$ is the $j^{th}$ bootstrap DM of length $B$;
\item Obtain $\hat{\bw}_{(j)} = \bD ^{-1} \hat{\breve{\bw}}_{(j)}$, where $\hat{\bw}_{(j)}$ is the high dimensional bootstrap DM of length $p$;
\item Repeat steps 1-3 $J$ times. Stack all $J$ values of $\hat{\bw}_{(j)}$ vertically and form $\hat{\bW} ^{*} = (
\hat{\bw}_{(1)}, \ldots, \hat{\bw}_{(J)} ) ^\intercal$, where $\hat{\bW} ^{*}$ is a $J \times p$ matrix. 
\end{enumerate}

Note the columns of $\hat{\bW} ^{*}$ are the bootstrap values of the DM corresponding to voxel $k$, from which we can form a distribution. There will be two types of distributions: unimodal and bimodal. The occurrence of bimodal distributions is due to the fact that the signs of the DM are not identifiable. Hence, we obtain voxel-wise p-values for $k \in \{1, \ldots, p \}$, by defining:
\begin{equation*} \label{eq:p-value}
P_k = 2 \mathbb{P} \big (t_{J-1} \geq \mid t_k\mid\big )
\end{equation*}
where $t_k = \min \bigg \{
\dfrac{\hat{\mu}_{k,1}} { \hat{\sigma}_{k,1}}, \dfrac{\hat{\mu}_{k,2}}{\hat{\sigma}_{k,2}}
\bigg\}$, $\hat{\mu}_{k,1}$ (resp. $\hat{\mu}_{k,2}$) and $\hat{\sigma}_{k,1}$ (resp. $\hat{\sigma}_{k,2}$) are the mean and standard deviation estimates of a mixed normal distribution. The \textit{mixtools} package \citep{mixtools} in \textit{R} includes EM-based procedures for estimating parameters from mixture distributions. 

\section{Simulation Study} \label{sec: sec5}

\subsection{Simulation Set-up}

Here we describe a simulation study to investigate the efficacy of our approach. Assume that, for every subject $i \in \{ 1, \ldots, n \}$, the mediator vector $\bM_i$ and the treatment $X_i$ can be jointly simulated from an independent, identically distributed multivariate normal distribution with known mean and variance.
 
In particular, let
\begin{equation} \label{eq:eq39}
\left.
\begin{array}{cc}
 \begin{pmatrix} \bM_i  \\
X_i \\
 \end{pmatrix} 
\end{array}
\right\vert \bmu, \bSigma
\sim
N_{p+1} \big( \bmu, \bSigma \big) 
\end{equation}
where  $\bmu = \begin{pmatrix}  \boldsymbol{\mu} ^M \\ \mu^X \end{pmatrix}$
and 
$\bSigma = \begin{pmatrix} 
\boldsymbol{\Sigma}^M & \boldsymbol{\Sigma}^{M,X} \\
\boldsymbol{\Sigma}^{X,M} & \Sigma^X 
\end{pmatrix}$.
Conditioning on $\bmu$ and $\bSigma$ we have
\begin{equation} \label{eq:eq_cond_M}
\big \{ \bM_i| X_i = x_i \big \} \sim N (\bar{\boldsymbol{\mu} }, \bar{ \boldsymbol{\Sigma} }),
\end{equation}
where $\bar{ \boldsymbol{\mu} } = \boldsymbol{\mu} ^M + \boldsymbol{\Sigma}^{M,X} [\Sigma^X] ^{ -1 } (x_i - \mu ^ X), $ and $\bar{ \boldsymbol{\Sigma} } = \boldsymbol{\Sigma} ^M - \boldsymbol{\Sigma} ^{M,X} [\Sigma ^X] ^{-1} \boldsymbol{ \Sigma} ^{X,M}$.
From \textcolor{blue}{ \eqref{eq:eq2} }:
\begin{equation*}
\E (\bM_i ^\intercal \bw_1 | X_i = x_i) = \alpha_0 + \alpha_1 x_i.
\end{equation*}
Solving \textcolor{blue}{ \eqref{eq:eq2} } and \textcolor{blue}{ \eqref{eq:eq_cond_M}}, we can write: 
\begin{eqnarray} \label{eq:eq42}
\alpha_0 &=& \bw_1[ \boldsymbol{\mu} ^M - \boldsymbol{\Sigma}^{M,X} [\Sigma^{X}]^{-1} \mu ^X ] ; \nonumber \\
\alpha_1 &=& \bw_1[ \boldsymbol{\Sigma}^{M,X} [\Sigma^{X}]^{-1} ] .
\end{eqnarray}
Moreover, 
\begin{eqnarray*}
\text{Var} (\bM_i \bw_1 | X_i = x_i) &=& \boldsymbol{\sigma}_\eta \\
&=& \bw_1^\intercal \textbf{Var} (\bM_i | X_i = x_i) \bw_1 \\
&=& \bw_1^\intercal \boldsymbol{\Sigma} ^M - \boldsymbol{\Sigma} ^{M,X} [\Sigma ^X] ^{-1} \boldsymbol{\Sigma} ^{X,M} \bw_1.
\end{eqnarray*}

\noindent Using these results we can outline the simulation process as follows: 
\begin{enumerate}
\item Set the values for $\bmu$ and $\bSigma$, and simulate $n$ pairs of $(\bM_i, X_i)$ according to \textcolor{blue}{ \eqref{eq:eq39} };
\item Set the values for $\beta_0, \beta_1,$ and $\gamma_1$, as well as $\bw_1$. Compute $\alpha_0$ and $\alpha_1$ using \textcolor{blue}{ \eqref{eq:eq42} }.
Consider these to be the true path coefficients $\btheta_1$ and the first direction of mediation $\bw_{1}$;
\item Simulate random error $\epsilon_i$ from a normal distribution with known mean and variance. Given $(\bM_i, X_i)$, $\epsilon_i$ , and the path coefficients, generate $Y_i$, $i = 1, \ldots n$, according to \textcolor{blue}{\eqref{eq:eq2}}. 
\end{enumerate}

\noindent The generated data $ \{ (X_i, Y_i, \bM_i ) \} _{i=1} ^n$ from Steps 1 and 3 are used as input in the LSEM. The outputs of the algorithm are compared with the true parameters.

Below we outline the four simulation studies that were performed. \\
\noindent \textbf{Simulation 1.} Let $p = 3$, $\bw_0 = (0.85, 0.17, 0.51)$, $\bmu = (2, 3, 4, 5)$, $\bSigma^{M, X} = (0.60, -0.90, 0.35)^\intercal$, and $\Sigma^X = 2.65$. Set the true path coefficients $(\beta_0, \beta_1, \gamma_1)$ equal to $(0.4, 0.2, 0.5)$. From \textcolor{blue}{ \eqref{eq:eq42} } it follows that $(\alpha_0, \alpha_1) = (3.23, 0.20)$. Assuming $\epsilon_i \sim N(0,1)$, we simulated $\{ X_i, Y_i, \bM_i \}_{i=1} ^n$, with $n = 10, 100, 500$, and $1,000$. Each set of simulations was repeated $1,000$ times, and the parameter estimates were recorded. 

\medskip
\noindent \textbf{Simulation 2.} Let $p = 10$, $\bw_0 = (0.42, 0.09, 0.25, 0.42, 0.17, 0.34, 0.51, 0.17, 0.17, 0.34)$, \\$\bmu = (2, 3, 4, 5, 4, 6, 2, 5, 8, 1, 3)$, $\bSigma^{M, X} = (-1.48, -0.51, -0.81,  0.98, -1.21,  0.53, -0.66,$ \\$ -0.73, -1.00,  0.29)^\intercal$, and $\Sigma^X = 5.10$.  Set the true pathway coefficients $(\beta_0, \beta_1, \gamma_1)$ to $(0.4, 0.2, 0.5)$. From \textcolor{blue}{ \eqref{eq:eq42} } it follows that  $(\alpha_0, \alpha_1) = (11.08, -0.20)$. Assuming $\epsilon_i \sim N(0,1)$, we simulated $\{ X_i, Y_i, \bM_i \}_{i=1} ^n$, with $n = 100$, and $1,000$. Each set of simulations was repeated $1,000$ times, and the parameter estimates were recorded. 

\medskip
\noindent \textbf{Simulation 3.} \label{itm: simulation3} Data are generated under the null hypothesis $\bw = 0$, i.e., $\bY$ is generated assuming no mediation effect. Consider $\bX$, a vector of length $1,149$, that ranges between $[44.3, 49.3]$ (both values chosen to mimic the fMRI data studied in the next section). Consider $(\beta_0, \gamma_1) = (-15, 0.5)$ and $\epsilon_i \sim N(0, 0.5)$. Generate $\bY_i$ according to \textcolor{blue}{ \eqref{eq:eq2} } with $\bw=0$, and let $M_i^{(j)} \sim N (m_i, s_i)$, where
$m_i \sim N (2, 5)$ and $s_i \sim N (20, 5)$. Here $M_i^{(j)}$ represents the simulated value of the $j^{th}$ voxel of trial $i$.
Using the technique introduced in Section \textcolor{blue} {\ref{sec: sec4}}, we obtain p-values for the estimated DM from the bootstrap distribution for each voxel. Fixing $\bX$, we independently generate $(\bW, \bY)$ 100 times, each time obtaining voxel-specific p-values. 

\medskip
\noindent \textbf{Simulation 4.} \label{itm: simulation4} Let $p=10,000$ and $n=1,000$.  First simulate $\bX$ from a truncated normal distribution $N^+(46.8, 2)$, truncated to take values in the range between $44.3$ and $49.3$.  Next construct $\bM$ under the assumption there are $1,000$ active and $9,000$ non-active voxels. This is achieved by simulating a vector of length $1,000$, corresponding to the active voxels, from a $N(1.5, 0.5)$ distribution, truncated to takes values in the range between $1$ and $2$.  These values were placed between two vectors of zeros each of length $4,500$, corresponding to non-active voxels, giving a vector of voxel-wise activity of length $10,000$.  Noise from a $N(0, 0.1)$ distribution was added to each voxel. This procedure was repeated for each of the $n$ subjects.  Entries of $\bw$ were set to weigh the voxels according to a Gaussian function, constrained to have norm $1$, centered at the middle voxel and designed to overlap in support with the $500$ centermost voxels. Finally, $\bY$ is simulated according to \textcolor{blue}{\eqref{eq:eq2:uni}}, where $(\beta_0,\gamma,\beta_1) = (-0.5,0.12,0.5)$ and $\eta_i \sim N(0,0.5)$.

\subsection{Simulation Results}

\begin{figure}[h]
\begin{subfigure}{.5\textwidth}
\centering
\includegraphics[width=.7\linewidth]{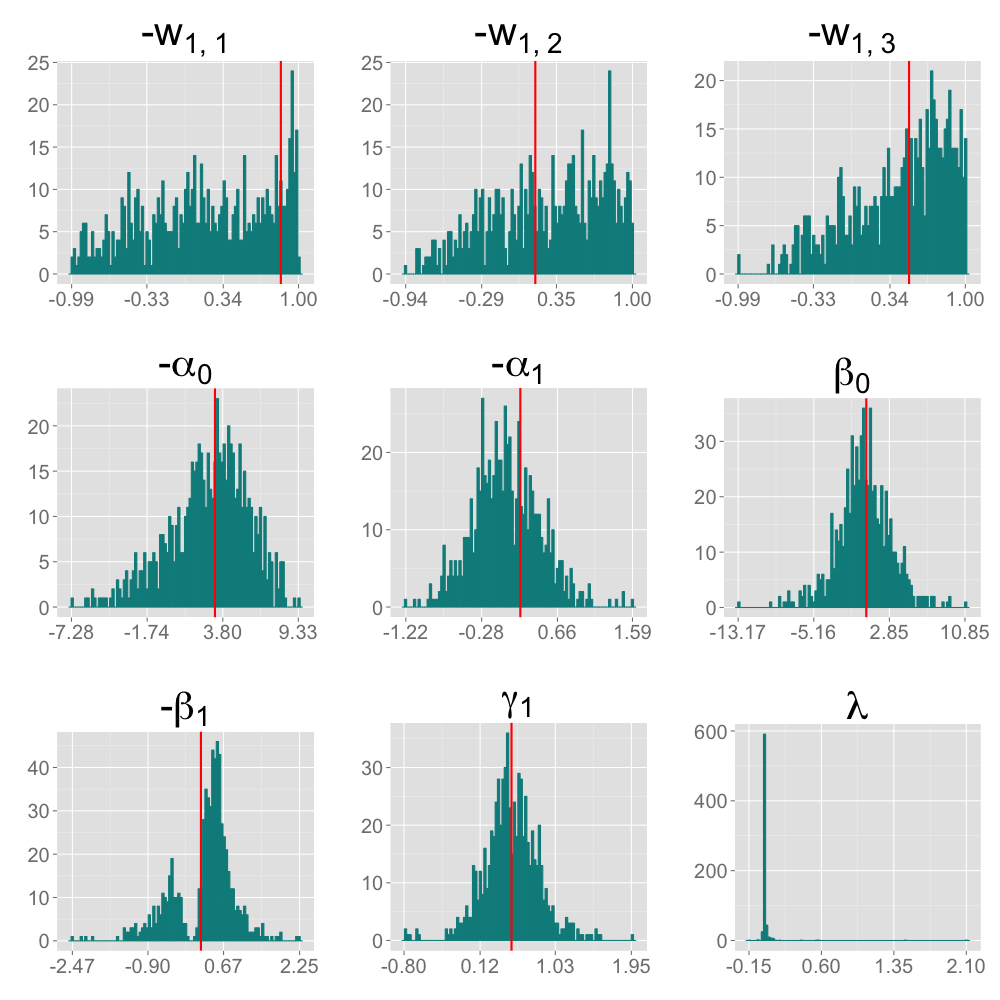}
\caption{10 by 3} 
\end{subfigure}%
\begin{subfigure}{.5\textwidth}
\centering
\includegraphics[width=.7\linewidth]{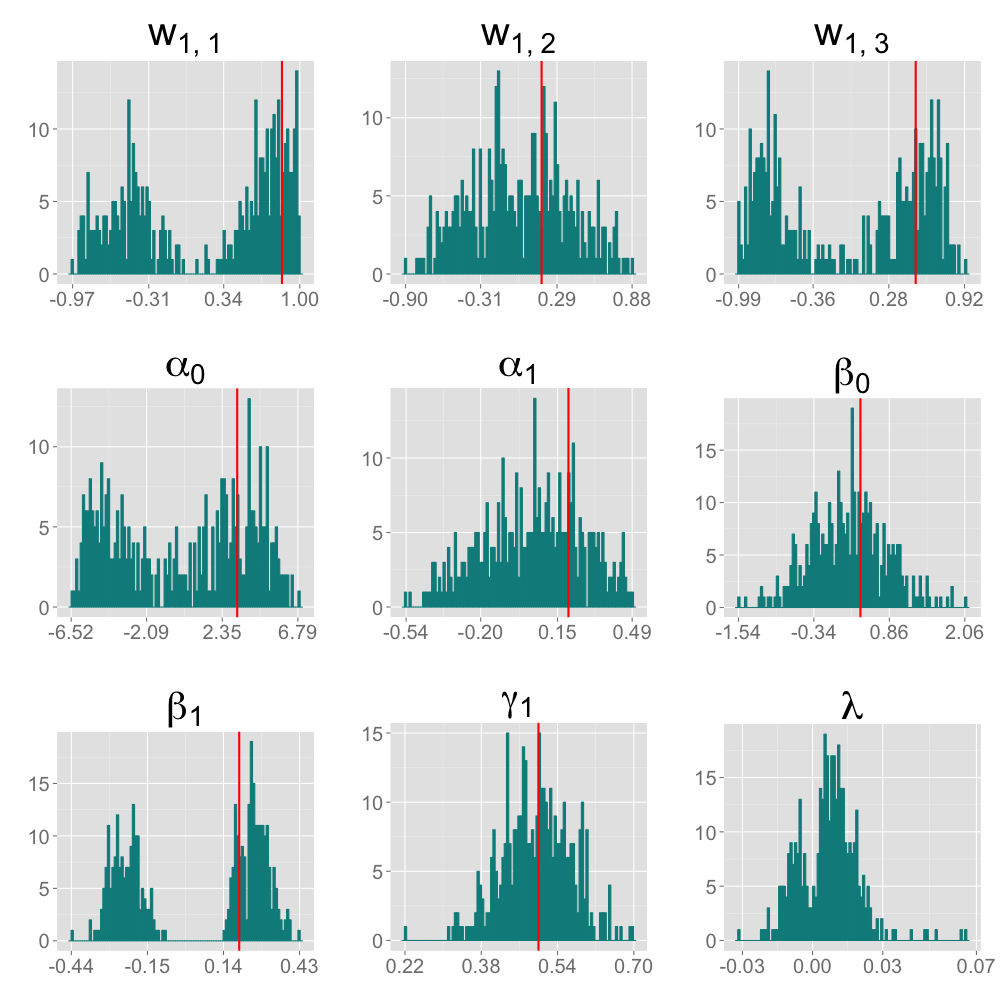}
\caption{100 by 3}
\end{subfigure}%

\begin{subfigure}{.5\textwidth}
\centering
\includegraphics[width=.7\linewidth]{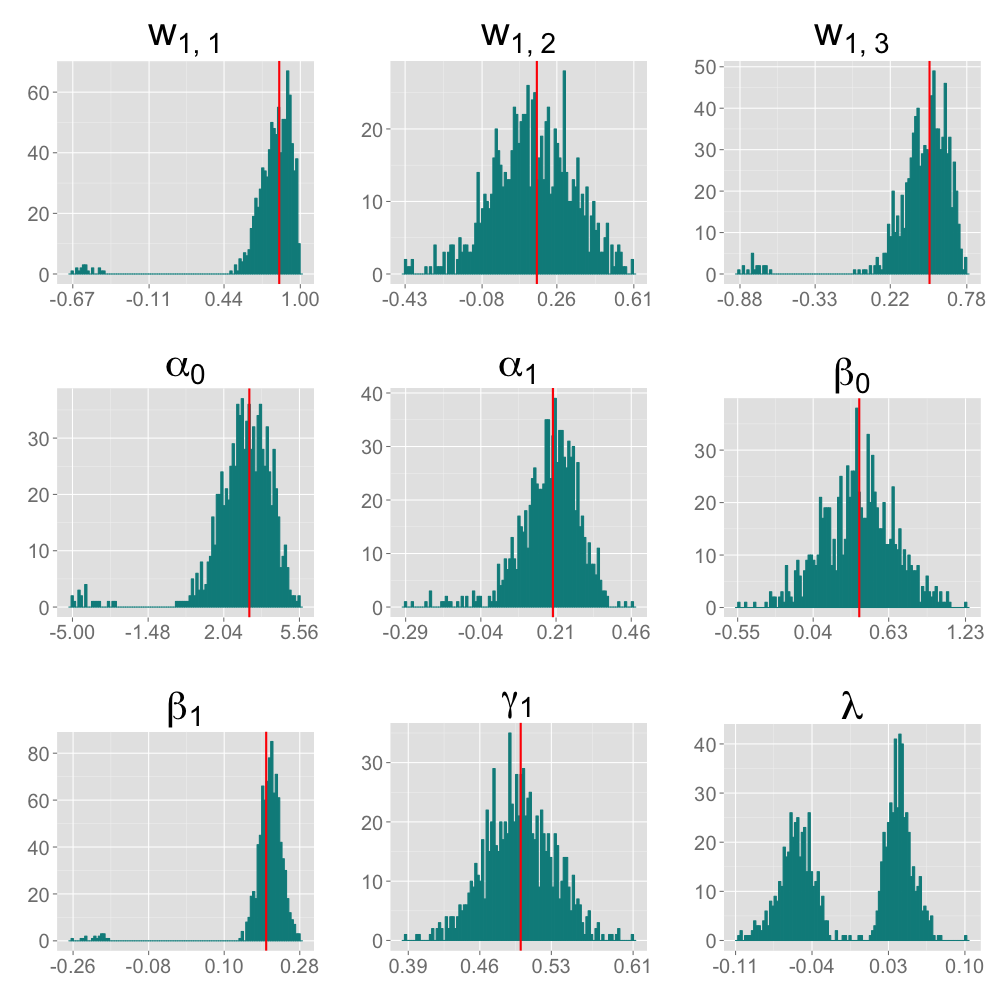}
\caption{500 by 3}
\end{subfigure}%
\begin{subfigure}{.5\textwidth}
\centering
\includegraphics[width=.7\linewidth]{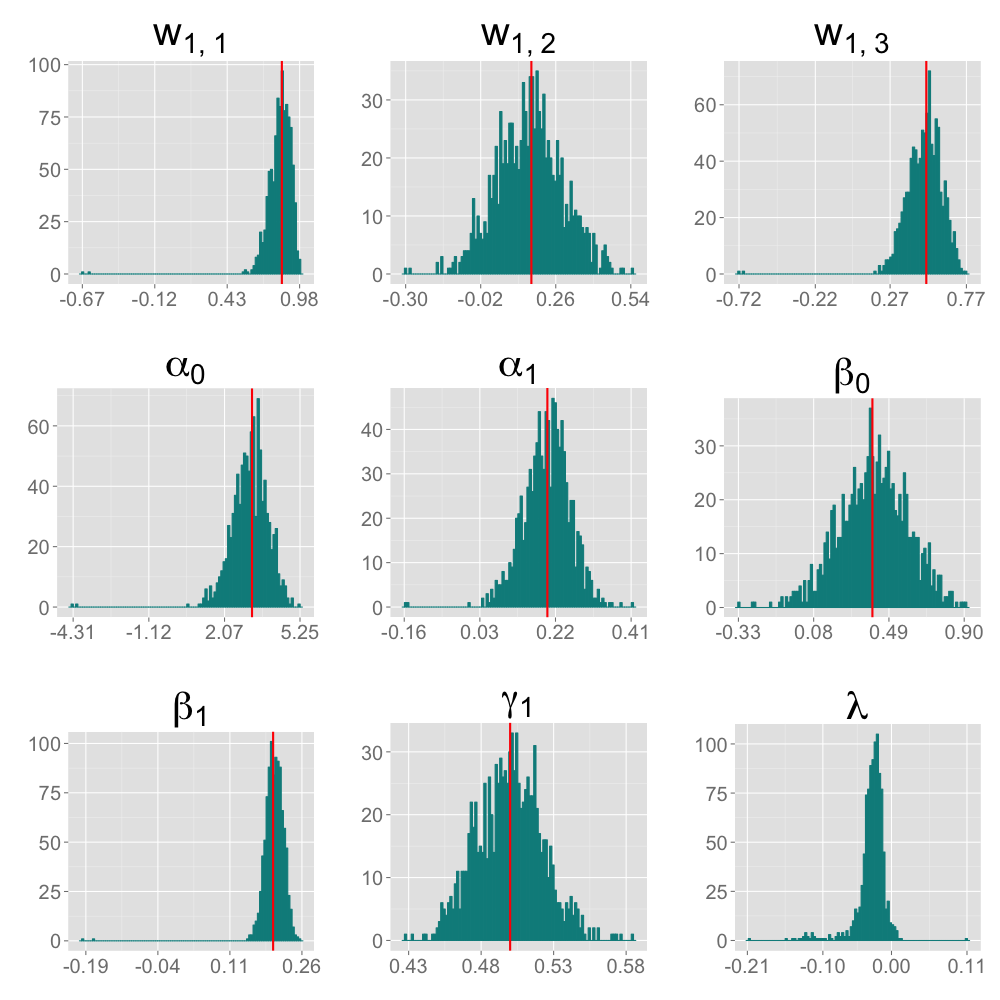}
\caption{1000 by 3}
\end{subfigure}%
\caption{\label{fig:fig2}
{\small Results for $p = 3$, when we increase sample size from 10 to 1,000 while keeping the ground truth values of $\bw$ and $\boldsymbol{\theta} = (\boldsymbol{\alpha}_0, \boldsymbol{\alpha}_1, \boldsymbol{\beta}_0, \boldsymbol{\beta}_1, \boldsymbol{\gamma}_1)$ fixed. Red lines indicate truth. }}
\end{figure}

\begin{figure}[h]
\begin{subfigure}{.5\textwidth}
\centering
\includegraphics[width=.9\linewidth]{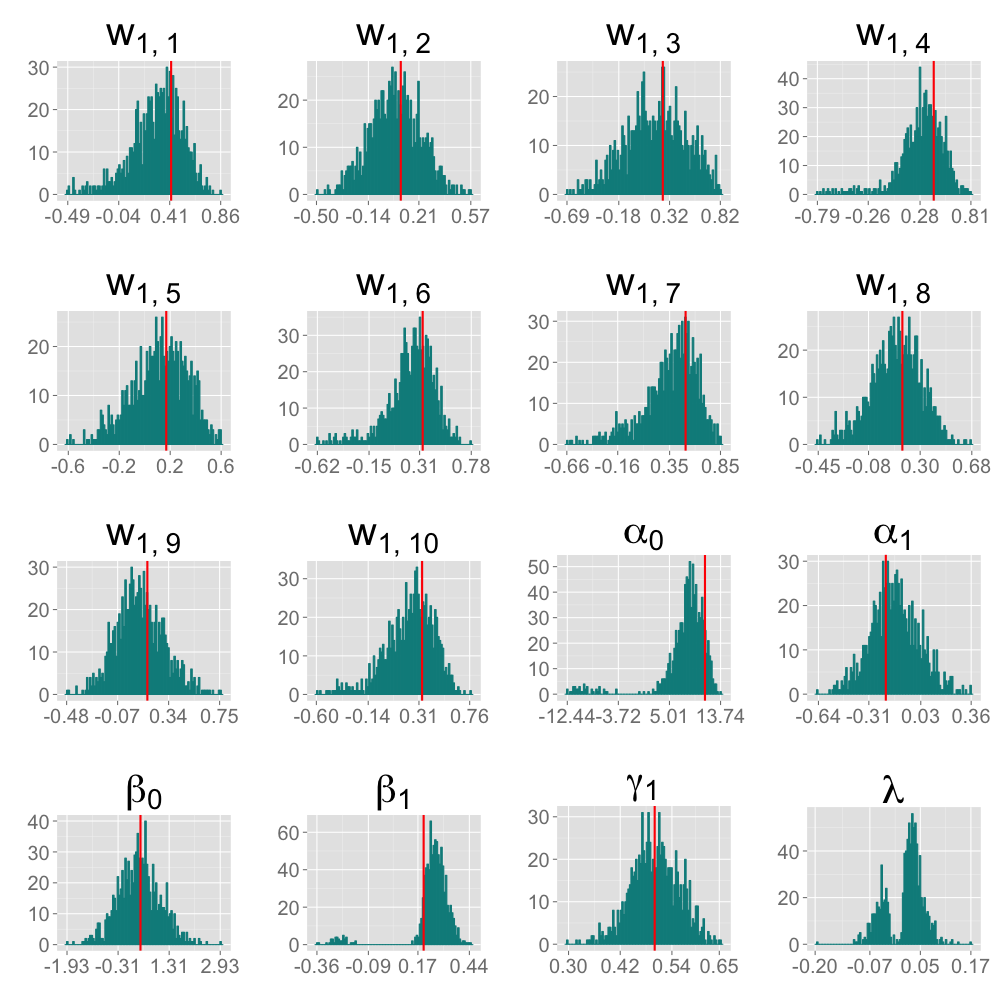}
\caption{100 by 10}
\end{subfigure}%
\begin{subfigure}{.5\textwidth}
\centering
\includegraphics[width=.9\linewidth]{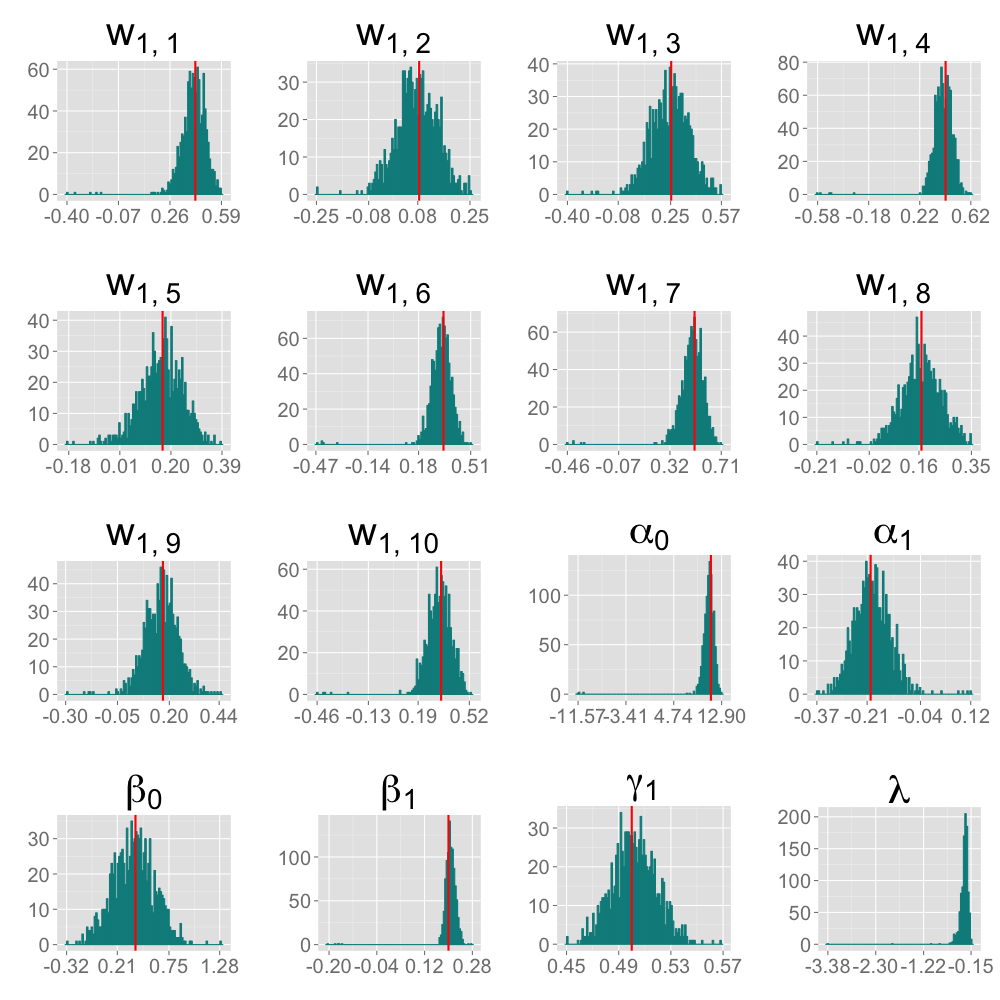}
\caption{1000 by 10}
\end{subfigure}%
\caption{\label{fig:fig3}
{\small Results for $p = 10$, when we increase sample size from 100 to 1,000 while keeping the ground truth values of $\bw$, and $\boldsymbol{\theta} = (\boldsymbol{\alpha}_0, \boldsymbol{\alpha}_1, \boldsymbol{\beta}_0, \boldsymbol{\beta}_1, \boldsymbol{\gamma}_1)$ fixed. Red lines indicate truth.}}
\end{figure}

Figures \textcolor{blue} {\ref{fig:fig2}} and \textcolor{blue} {\ref{fig:fig3}} show the results of Simulations 1 and 2. Figure \textcolor{blue} {\ref{fig:fig2}} a-d display results for the case when $p=3$, and the sample size is $10$, $100$, $500$, and $1,000$, respectively.  Figure \textcolor{blue} {\ref{fig:fig3}} a-b display results for $p=10$, and the sample size is $100$ and $1,000$.  As the sample size increases, the estimates become more accurate, while the distribution becomes increasingly normal with a smaller standard deviation. The sign of the estimator is difficult to determine for smaller samples sizes, but becomes more consistent as the sample size increases. 

\begin{center}
\begin{tabular}{|c|c|c|c|} 
\cline{3-4} 
\multicolumn{1}{c}{} & & \multicolumn{2}{c|}{p}\tabularnewline
\cline{3-4} 
\multicolumn{1}{c}{} & & 3 & 10\tabularnewline
\hline 
\multirow{5}{*}{n} & 10 & 694 & ---\tabularnewline
\cline{2-4} 
& 100 & 387 & 923\tabularnewline
\cline{2-4} 
& 300 & 633 & 984\tabularnewline
\cline{2-4} 
& 500 & 897 & 1,000\tabularnewline
\cline{2-4} 
& 1,000 & 1,000 & 1,000\tabularnewline
\hline 
\end{tabular} 
\captionof{table}{The turn-out rate for different $n$ and $p$ combinations per 1,000 Simulations \label{tab:tab1} } 
\end{center}

Moreover, for fixed $p$, the turn-out rate (the number of estimating results an algorithm produces out of a fixed number of simulations) increases with $n$; see Table I. 
For fixed $n$, the turn-out rate improves with increasing $p$.  The reason why some runs do not produce a result is that the function $\lambda (\boldsymbol{\theta}) $ is not well behaved in small sample sizes, and the Newton-Raphson optimization algorithm fails at one of the intermediary steps. When $p$ is sufficiently large or high dimensional, the algorithm seems to improve. If $p \sim 3$, the algorithm runs better when we have sufficiently large sample size (e.g., $n \sim 300$). Performance of the algorithm improves with more refined grid points, but this comes at the expense of computational efficiency.

\begin{figure}[htbp] 
\begin{center}
\includegraphics[width=10cm]{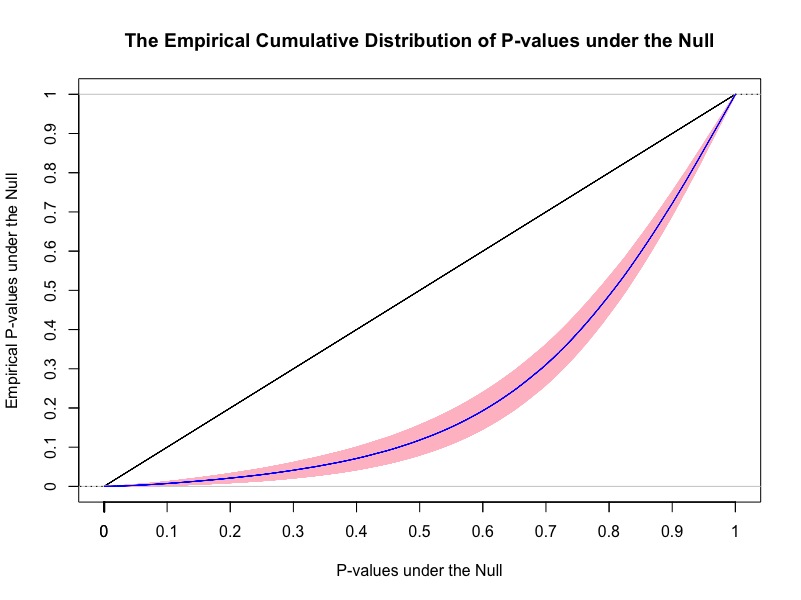}\\
\caption{
{ \footnotesize
The empirical p-value plotted against the theoretical p-value. The straight line indicates exact correspondence between the two, and $95 \%$ confidence bands are shown in pink. }}
\label{fig:Voxel_wise_p_values}
\end{center}
\end{figure}

\begin{figure}[htbp] 
\begin{center} 
\includegraphics[width=10cm]{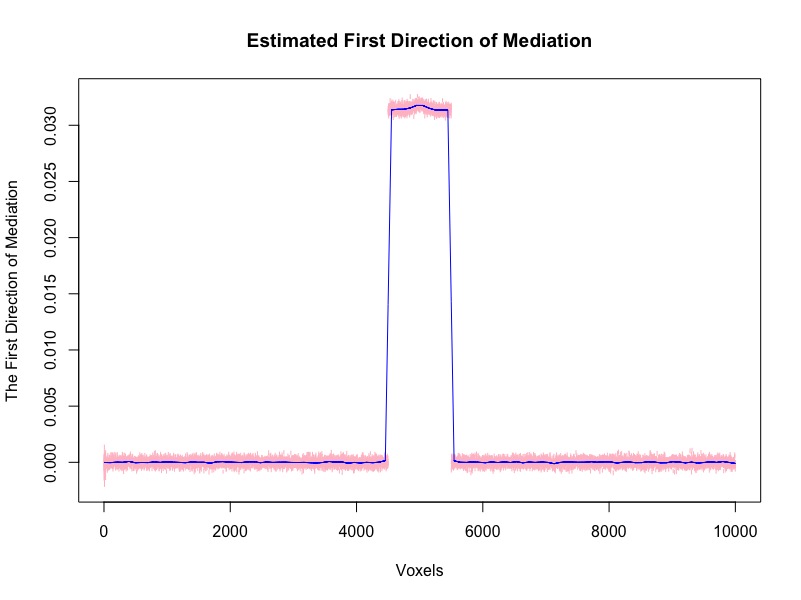}\\
\caption{
{ \footnotesize
Bootstrap confidence bands of the estimated first direction of mediation computed from $100$ bootstrap repetitions. The simulated mediator is designed to show strong activations in the center $1,000$ voxels. The estimated first direction of mediation is consistent with the simulated signals (blue line). }}
\label{fig:estimated_band}
\end{center}
\end{figure}

The results of Simulation 3 are shown in Figure \textcolor{blue} {\ref{fig:Voxel_wise_p_values}}. Here the empirical p-values under the null, represented by the portion of voxels that fall below a certain threshold, are plotted against the theoretical p-values. $95 \%$ confidence bounds are shown in pink. Clearly, the approach provides adequate control of the false positive rate in the null setting, albeit with somewhat over-conservative results.
Finally, Fig. \textcolor{red}{\ref{fig:estimated_band}} shows bootstrap confidence bands for the estimated first direction of mediation from 100 bootstrap repetitions. Recall that the mediator is designed to to have $1,000$ active voxels. Clearly, the estimated first direction of mediation is consistent with the simulated signal.

\section{An fMRI Study of Thermal Pain} \label{sec: sec6}

\subsection{Data Description} 

The data comes from the fMRI study of thermal pain described in the Introduction. 
A total of $33$ healthy, right-handed participants completed the study (age $27.9 \pm 9.0$ years, $22$ females). All participants provided informed consent, and the Columbia University Institutional Review Board approved the study. 

The experiment consisted of a total of nine runs. Seven runs were ``passive'', in which participants passively experienced and rated the heat stimuli, and two runs were ``regulation'', where the participants imagined the stimuli to be more or less painful than they actually were, in one run each (counterbalanced in order across participants). In this paper we consider only the seven passive runs, consisting of between $58 - 75$ separate trials (thermal stimulation repetitions). During each trial, thermal stimulations were delivered to the volar surface of the left inner forearm. Each stimulus lasted $12.5$s, with $3$s ramp-up and $2$s ramp-down periods and $7.5$s at the target temperature. Six levels of temperature, ranging from $44.3 - 49.3$ $\degree$C in increments of $1$ $\degree$C, were administered to each participant. Each stimulus was followed by a $4.5 - 8.5$s long pre-rating period, after which participants rated the intensity of the pain on a scale of $0$ to $100$. Each trial concluded with a $5 - 9$s resting period.

Whole-brain fMRI data was acquired on a 3T Philips Achieva TX scanner at Columbia University. Structural images were acquired using high-resolution T1 spoiled gradient recall (SPGR) images  with the intention of using them for anatomical localization and warping to a standard space. Functional EPI images were acquired with TR = $2000$ms, TE = $20$ms, field of view = $224$mm, $64 \times 64$ matrix, $3 \times 3 \times 3$mm$^3$ voxels, $42$ interleaved slices, parallel imaging, SENSE factor $1.5$. For each subject, structural images were co-registered to the mean functional image using the iterative mutual information-based algorithm implemented in SPM8\footnote{http://www.fil.ion.ucl.ac.uk/spm/}. Subsequently, structural images were normalized to MNI space using SPM8's generative segment-and-normalize algorithm. Prior to preprocessing of functional images, the first four volumes were removed to allow for image intensity stabilization. Outliers were identified using the Mahalanobis distance for the matrix of slice-wise mean and the standard deviation values. The functional images were corrected for differences in slice-timing, and were motion corrected using SPM8. The functional images were warped to SPMs normative atlas using warping parameters estimated from coregistered, high resolution structural images, and smoothed with an $8$mm FWHM Gaussian kernel. A high-pass filter of $180$s was applied to the time series data. 

A single trial analysis approach was used, by constructing a general linear model (GLM) design matrix with separate regressors for each trial (\cite{rissman2004measuring}; \cite{mumford2012deconvolving}). Boxcar regressors, convolved with the canonical hemodynamic response function, were constructed to model periods for the thermal stimulation and rating periods for each trial. Other regressors that were not of direct interest included (a) intercepts for each run; (b) linear drift across time within each run; (c) the six estimated head movement parameters ($x$, $y$, $z$, roll, pitch, and yaw), their mean-centered squares, derivatives, and squared derivative for each run; (d) indicator vectors for outlier time points; (e) indicator vectors for the first two images in each run; (f) signal from white matter and ventricles. Using the results of the GLM analysis, whole-brain maps of activation were computed.

In summary, $X_{ij}$ and $Y_{ij}$ are the temperature level and pain rating, respectively, assigned on trial $j$ to subject $i$, and ${\bM}_{ij} = (M_{ij}^{(1)}, M_{ij}^{(2)}, \ldots M_{ij}^{(p)} )^\intercal \in \mathbb{R}^p$ is the whole-brain activation measured over $p = 206, 777$ voxels, defined as the regression parameter corresponding to the stimulus in the associated GLM. In addition, $i \in \{ 1,\dots, I\}$ and $j \in \{ 1,\dots, J_i\} $, where $I=33$ and $J_i$ takes subject-specific values between $58-75$. The data was arranged in a matrix $\bM$ of dimension $1,149 \times 206,777$, where each row consists of activation from a single trial on a single subject over $206,777$ voxels, and each column is voxel-specific. The temperature level and reported pain are represented as the vectors $\bx$ and $\by$, respectively, both of length $1,149$.

\subsection{Results} 
Each DM corresponding to $\Delta = (\bx, \by, \bM)$, is a vector of length $206,777$, whose estimation is computationally infeasible without first performing data reduction. Hence, we use the GPVD approach outlined in Section \textcolor{blue} {\ref{sec: sec3.5}}.  We choose $\tilde{\bw}$ to have dimension $B=35$, to ensure that the number of rows of $\bD$ is less than or equal to the minimum number of trials per subject. This value ensures that $80 \%$ of the total variability of $\bM$ is explained after dimension reduction. 
The population-specific matrix $\bD$ of dimension $35 \times 206,777$ was obtained according to \textcolor{blue} {\eqref{eq:eq29}}, and the lower dimensional mediation matrix $\tilde{\bM}$ of dimension $1,149 \times 35$, according to \textcolor{blue} {\eqref{eq:M_tilde}}. The terms $(\bx, \by, \tilde{\bM})$ were placed into the algorithm outlined in \textcolor{blue} {\eqref{eq:eq15}} - \textcolor{blue} {\eqref{eq:eq17}}, using starting values $\btheta_1^{(0)} = 0.1 \times \bJ_5$, and $\bw_1^{(0)} = 0.1 \times \bJ_{35}$. Finally,  $\hat{\bw}$, of length $206,777$, was computed using \textcolor{blue} {\eqref{eq: ginverse}}.

\begin{figure}[htbp]
\begin{center}  
\includegraphics[width=17 cm]{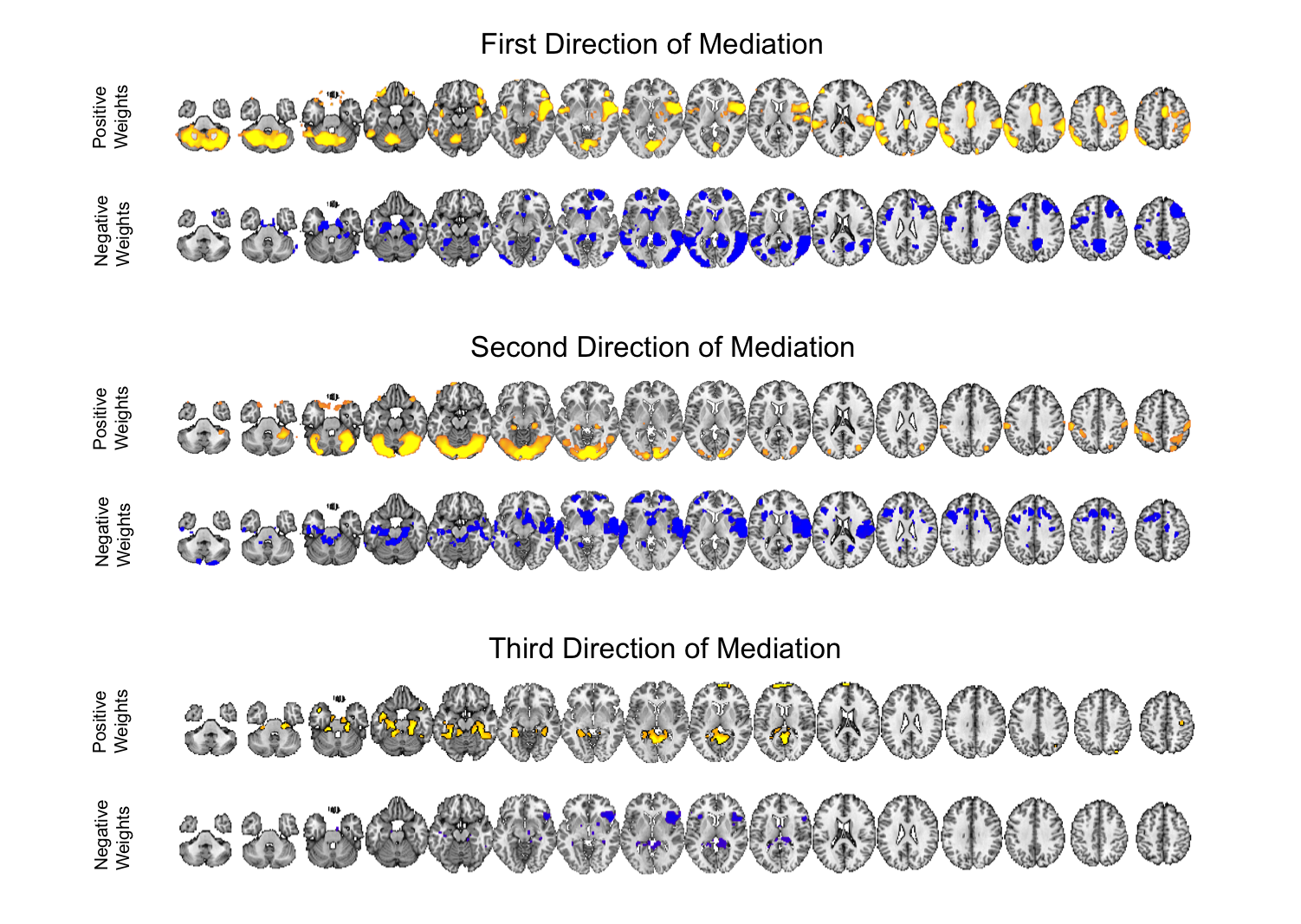}
\caption{
{ \footnotesize
Weight maps for the first three Directions of Mediation fit using data from the fMRI study of thermal pain.
Significant weights are separated into those with positive and negative values, respectively, for the each DM.
All maps are thresholded using FDR correction with $q=0.05$. }}
\label{fig:Results}
\end{center}
\end{figure}

We compute the first three DMs and obtained estimates of $ \hat{\boldsymbol { \theta}}_1 = (-3769.30, 96.32, -13.86, \\0.00075, 0.40)$, $\hat{\boldsymbol { \theta}}_2  = (-695.85, -24.11, -13.86, 0.00075,   -1.06 \times 10 ^{-7}, 0.40)$, and\\
$\hat{\boldsymbol { \theta}}_3  = (1.35,  -0.03, -13.86, 0.00075, -3.585 \times 10^{-7},  -5.5 \times 10^{-9}, 0.40)$. Figure \textcolor{blue} {\ref{fig:Results}} shows the weight maps for the first three Directions of Mediation, thresholded using FDR correction with $q=0.05$, separated according to whether the weight values were positive or negative. 

The map is consistent with regions typically considered active in pain research, but also reveals some interesting structure that has not been uncovered by previous methods. The first direction of mediation shows positive weights on both targets of ascending nociceptive (pain-related) pathways, including the anterior cingulate, mid-insula, posterior insula, parietal operculum/S2, the approximate hand area of S1, and cerebellum.  Negative weights were found in areas often anti-correlated with pain, including parts of the lateral prefrontal cortex, parahippocampal cortex, and ventral caudate, and other regions including anterior frontal cortex, temporal cortex, and precuneus.  These are associated with distinct classes of functions other than physical pain and are not thought to contain nociceptive neurons, but are still thought to play a role in mediating pain by processing elements of the context in which the pain occurs.

The second direction of mediation is interesting because it also contains some nociceptive targets and other, non-nociceptive regions that partially overlap with and are partially distinct from the first direction.  This component splits nociceptive regions, with positive weights on S1 and negative weights on the parietal operculum/S2 and amygdala, possibly revealing dynamics of variation among pain processing regions once the first direction of mediation is accounted for.  Positive weights are found on visual and superior cerebellar regions and parts of the hippocampus, and negative weights on the nucleus accumbens/ventral striatum and parts of dorsolateral and superior prefrontal cortex. The latter often correlate negatively with pain.  

Finally, the third direction of mediation involves parahippocampal cortex and anterior insula/VLPFC, both regions related to pain.


\section{Discussion} \label{sec:Discussion}

This paper addresses the problem of mediation analysis in the high-dimensional setting.  The first DM is the linear combination of the elements of a vector of potential mediators that maximizes the likelihood of the underlying three variable SEM. Subsequent directions can be found that maximize the likelihood of the SEM conditional on being orthogonal to previous directions. 

The causal interpretation for the parameters of the DM approach rests on a strong untestable assumption, namely sequential ignorability. For example, the assumption $Y(x,m) \independent {\bM} | X$ would be valid if the mediators were randomly assigned to the subjects. However, this is not the case here, and instead, we must assume that they behave as if they were. This assumption is unverifiable in practice and ultimately depends on context. In the neuroimaging setting, its validity may differ across brain regions, making causal claims more difficult to access. That said, we believe the proposed approach still has utility for performing exploratory mediation analysis and detecting sets of regions that potentially mediate the relationship between treatment and outcome, allowing these regions to be explored further in more targeted studies. 

It should further be noted that when deriving the direct and indirect effect in section \ref{MCMM} we assumed each subject was scanned under one condition. However, in most fMRI experiments subjects are scanned under multiple conditions, as in our motivating pain data set. Extension of the casual model to this case will allow for single subject studies of mediation in which unit direct effects on the mediators and unit total effects on outcomes are observed. In some instances, the observability of these unit effects can be used to estimate both single subject and population averaged models under weaker and/or alternative conditions than those in \ref{eq:mediation}. We leave this extension for future work.
In addition, in our motivating example the mediator is brain activation measured with error. Thus, an extension would be to modify the model to deal with systematic errors of measurement in the mediating variable (\cite{sobel2014causal}).

One property of the DM framework is that the signs of the estimates are unidentifiable. To address this issue, there are two possible solutions. First, we can use Bayesian methods to apply a sign constraint based on prior knowledge. Second, if the magnitude of the voxel-wise mediation effect is of interest, we can consider a non-negativity constraint. For example, through re-parameterization, as by setting $w=\exp(v)$. This can be necessary because, under some circumstances, the coexistence of positive and negative elements of $\bw$ could cancel out potential mediation effects. For example, assume $\bM = (0.5, 0.4, 0.9)$ and $\bw = ( 0.577, 0.577, - 0.577)^\intercal$. Then $\bM \bw = 0$, making the estimate of $\beta_1$ unavailable. It, however, does not necessarily imply the non-existence of a mediation effect. 

In many settings, the response $\bY$ and the mediator $\bM$ are not necessarily normally distributed, but instead follow some distribution from the exponential family. It can be shown that we can estimate both the DMs and path coefficients under this setting using a GEE-like method. Essentially, conditioning on the DM, the path coefficient can be estimated using two sets of GEEs. The DM can then be estimated conditioning on the estimated coefficients.

\section*{Acknowledgement}
This research was partially supported by NIH grants R01EB016061, R01DA035484 and P41 EB015909, and NSF grant 0631637. The authors would like to thank Tianchen Qian of Johns Hopkins Bloomberg School of Public Health (JHSPH) for his insightful comments on deriving the asymptotic property of the estimates, and Stephen Cristiano, Bin He, Haoyu Zhang, and Shen Xu of JHSPH for their valuable suggestions.  

\newpage
\bibliographystyle{apalike}
\bibliography{PDM.bib}

\end{document}